\documentclass{article}
\usepackage[english]{babel}
\usepackage[utf8]{inputenc}
\usepackage{johd}
\usepackage{amsmath}
\usepackage{siunitx}
\usepackage{rotating}
\usepackage{enumitem}
\usepackage{orcidlink}
\usepackage{listings}
\title{Making the complete OpenAIRE citation graph easily accessible through compact data representation}

\author{Joakim Skarding$^{a}$\orcidlink{0000-0001-8509-658X}, Pavel Sanda$^{a}$\orcidlink{0000-0002-2554-0180}\\
        \small $^{a}$Institute of Computer Science of the Czech Academy of Sciences, Prague, Czech Republic \\
        \small $^{*}$Corresponding author: Pavel Sanda; \tt{sanda@cs.cas.cz} \\
}
\date{}

\begin{document}
\maketitle

\noindent{Data paper}

\begin{abstract} 
\noindent The OpenAIRE graph contains a large citation graph dataset, with over 200 million publications and over 2 billion citations. The current graph is available as a dump with metadata which, when uncompressed, totals $\sim 2.5$\,TB. This makes it hard to process on conventional computers. To make this network more accessible for the community, we provide a processed OpenAIRE graph which is downscaled to 16\,GB RAM, while preserving the full graph structure. Apart from this we offer the processed data in a very simple format, which allows for further straightforward manipulation. We also provide (1) a Python pipeline, which can be used to process the next releases of the OpenAIRE graph, and (2) a larger version of the dataset including more publication fields such as, the title, list of authors.
\end{abstract}

\noindent\\

\section{Overview}
%Indicate where the dataset repository is located – a DOI of the openly available dataset being described is required.
\paragraph{Repository location}
We use two repositories, Zenodo to share our data and fixed version of the processing pipeline, and git repository on Codeberg to facilitate sharing and further development of the processing pipeline.
\begin{description}
    \item[Data repository] \href{https://zenodo.org/records/19207803}{https://zenodo.org/records/19207803}
    \item[Code repository] \href{https://codeberg.org/Zmeos/OpenAIRE-citation-extraction}{https://codeberg.org/Zmeos/OpenAIRE-citation-extraction}
\end{description}

\paragraph{Context} 
The OpenAIRE graph \citep{rettberg2012openaire, manghi_2025_17098012} is a large knowledge graph storing several kinds of research data. In this work we focus on extracting the citations and publications from that dataset, distilling the information to make it more accessible.

There are several other citations graphs available -- OpenAlex \citep{priem2022openalex}, Open Research Knowledge Graph \citep{jaradeh2019open}, Crossref \citep{hendricks2020crossref} and OpenCitations \citep{peroni2020opencitations}. These datasets differ in many ways, some are only citation graphs, while others like OpenAlex and OpenAIRE are research knowledge graphs that record much more than just citations and publications. The datasets also differ in their publication coverage \citep{martin2021google, culbert2025reference} and how the works are indexed, e.g. how they determine taxonomy \citep{ciuciu2025scientific}. Details of publication coverage between OpenAIRE and other open datasets are not well explored -- a subset of OpenAIRE has been compared to OpenAlex~\citep{ciuciu2024assessing}, but to the best of our knowledge, no coverage analysis to the extent of previous works like \citep{martin2021google} or \citep{culbert2025reference} has taken place. It is not sufficient to compare publication numbers since these datasets aggregate publications from multiple sources and a large node number might simply be due to poor de-duplication.

The OpenAIRE initiative offers several APIs, a search engine,  cloud access through Google BigQuery, and a complete dataset dump \citep{manghi_2025_17098012}. The API and search engine serve as accessible avenues for any researcher, however, they are limited in the scope of data that can be efficiently accessed. A scan of the complete graph is outside the free tier of BigQuery, at the time of writing. If a researcher is interested in large portions of the graph, or the graph as a whole, then the dump is more suitable. 

Processing the whole graph dump requires software coding skills, a large amount of memory, storage and computational power, resources not readily accessible to many scholars in the humanities \citep{dederke2024representation}. Accessing the raw data also requires solid understanding of the OpenAIRE JSON schema. To fill this gap, we pre-process the dump, distilling it into a more manageable size, and distributing it as simply structured plain text and Parquet files, making the whole graph, more accessible.
 Similar work has also been done for OpenAlex \citep{illinoisdatabankIDB-7362697}.

%The data was processed as part of an interdisciplinary project to explore spread of scientific knowledge across history. 

%The authors of this paper are not affiliated with OpenAIRE, and do not claim to be the owners of the OpenAIRE graph, we have only preprocessed it, to make it more accessible.

% Start of potential paragraph on other citation graph and why making OpenAIRE available is important:
%There are other global citations graphs OpenAlex \cite{priem2022openalex}, OpenAIRE \cite{rettberg2012openaire} \cite{manghi_2025_17098012}, Open Research Knowledge Graph \cite{jaradeh2019open}, Crossref \cite{hendricks2020crossref}

%Was this data produced as part of a research project, thesis, course work, or is this data used in a paper(s)? If so, please list the appropriate bibliographic information here. Note: This journal uses a style based on the APA system (see \href{https://openhumanitiesdata.metajnl.com/about/submissions/#References}{here}).\\

\section{Method}
The OpenAIRE graph consists of many entities and many relations. These are stored in many files in the dump. We only need the publication files (nodes) and the relation files (citations/edges) to extract the citation network. The publication files themselves contain a lot of information about publications, only some of which is relevant to most researchers. The relation files include many types of relations. For this dataset we are only interested in citations (the type "Cites"). The following list shows the generic steps we followed to extract the citation network and minimize memory footprint. The list is organized by python files and shows the file responsible for each step. (the full code is available in the Codeberg repository).
\paragraph{Steps} %The series of procedures followed to produce the dataset. This should include any source data used, as well as software and instrumentation involved.
\begin{enumerate}[start=0]
    \item \texttt{step0\_download\_data\_and\_extract.py}\\ Download the complete OpenAIRE graph dump \citep{manghi_2025_17098012}, including all publication and relation files. When calling the pipeline, this is an optional step.
    \item \texttt{step1\_extract\_raw.py}\\ Partially extract the compressed dump files.
    \item \texttt{step2\_publications.py} 
    \begin{enumerate}
        \item Filters records, obtaining publication-type only. 
        \item Flatten nested JSON structures to obtain a tabular representation.
        \item Generate new, more memory-efficient(\texttt{int32}) nodeIDs, and build a hashtable for translating between OpenAIRE IDs and our nodeIDs. 
        \item For the TSV files; replace any tabs or newlines with whitespace
        \item Export TSV and Parquet publication files
    \end{enumerate} 
    \item \texttt{step3\_citations.py} 
    \begin{enumerate}
        \item Process relation files using PySpark \citep{zaharia2016apache}, retaining only relations of type \texttt{Cites} and extracting source and target identifiers. 
        \item Use hashtable produced by \texttt{step2\_publications.py} to replace OpenAIRE IDs. Update all citation relations to reference the new identifiers.
        \item Export TSV and Parquet citation files
    \end{enumerate}
    \item \texttt{step4\_distribution.py}\\ Compresses the TSV files to \texttt{xz}.
\end{enumerate}

This achieves a distilled dataset where the relations are efficiently stored using pairs of short integers.

\paragraph{Quality control} 
Several validation scripts are run after processing, to control for mistakes in the processing. The scripts check whether any entries from the original dump were lost and report any missing values.
%The script counts the number of publications and relations in the original dump, and compares to the distilled dump.
\begin{table}[h]
\centering
\begin{tabular}{lll}
\hline
Script & Metric & Value \\
\hline

\multicolumn{3}{l}{\textbf{full\_id\_coverage (pass)}} \\
 & raw\_count     & 205841448 \\
 & parquet\_count & 205841448 \\
 & missing\_count & 0 \\
 & extra\_count   & 0 \\

\hline
\multicolumn{3}{l}{\textbf{distinct\_constraints (pass)}} \\
 & publications total     & 205841448 \\
 & publications distinct  & 205841448 \\
 & publications duplicates& 0 \\
 & citations total        & 2184347684 \\
 & citations distinct     & 2184347684 \\
 & citations duplicates   & 0 \\
 & citations null\_rows   & 0 \\

\hline
\multicolumn{3}{l}{\textbf{format\_checks (pass)}} \\
 & nodeid gaps           & 0 \\
 & nodeid contiguous     & true \\

\hline
\end{tabular}
\caption{Combined validation script output. This is the automated output log of the validation scripts.}
\label{tab:validation}
\end{table}

 \begin{itemize}
      \item \texttt{full\_id\_coverage.py} Verifies that every publication in the raw source data is present in the final output.
      \item \texttt{distinct\_constraints.py} Checks that there are no duplicate publications or duplicate citation edges in the output.
      \item \texttt{format\_checks.py} Verifies that publication node IDs form a complete, gap-free sequence starting from zero.
  \end{itemize}  
  Additionally the file \texttt{run\_all\_validations.py} runs all validation scripts, collects their results, and writes a consolidated summary report. The entire validation pipeline is automatically run after the processing pipeline completes.

\paragraph{Output} The processing pipeline takes the full OpenAIRE dump as input and transforms it into the TSV (tab-separated values) and equivalent Parquet files described below. TSV is a variant of a more common CSV. We chose TSV  because several of the text fields in the publications\_large.tsv routinely contained commas, and our attempts at quoting the commas turned out to be error prone. Parquet is a binary format for high-performance data processing.

\begin{itemize}
    \item \verb|citations.tsv.xz| and \verb|citations.parquet| -- a simple edge list of the graph (all the citations). Each row has a simple form of two connected nodeIDs, e.g.: \verb|159486578   118392581|
    \item \verb|publications.tsv.xz| and \verb|publications.parquet| -- all graph nodes (publications), it contains only the nodeID and, when available, the DOI. Each row has a simple form, e.g.:
    \verb|14209 10.3931/e-rara-45685|
    \item \verb|publications_large.tsv.xz| and \verb|publications_large.parquet|-- includes the same number of nodes as the publication files, but includes additional fields, e.g. title, authors, description, etc. For a full overview of fields, see Table \ref{tab:column_sizes_python_types_desc_left}.
    \item \verb|pipeline.tar.xz| -- processing pipeline which transform full OpenAIRE dump into the dataset files above. Full details are documented in included \verb|README.md|. The source code is also present online at \\
    \href{https://codeberg.org/Zmeos/OpenAIRE-citation-extraction}{https://codeberg.org/Zmeos/OpenAIRE-citation-extraction}.
\end{itemize}

The \texttt{pid\_*} columns (included in publications\_large, see Table \ref{tab:column_sizes_python_types_desc_left}) are only extracted schemes from the \texttt{pids} field in the dump which OpenAIRE populates with identifiers collected from authoritative source. The instances field (a nested field of additional metadata, Table \ref{tab:filtered_fields}) also includes identifiers, those were not collected, with the exception of the \texttt{primary\_doi} field in the publications file, there the doi is drawn from the identifiers field if it is not found in the pids field.

\paragraph{Memory requirements}
Below, we summarize the hardware requirements for the transformed dataset and the original OpenAire full dataset dump. For loading the citations.tsv into memory with \verb|int32| using the Pandas library in Python, the following can be used: %\\\verb|df = pd.read_csv("citations.tsv.xz", sep="\t", dtype={"source_nodeId": "int32", "target_nodeId": "int32"})|
\begin{lstlisting}
df_refs = pd.read_csv(
    "citations.tsv.xz", sep="\t",
    dtype={"source_nodeId": "int32", "target_nodeId": "int32"}
)
\end{lstlisting}
The data formats (TSV and Parquet) are of course generic and can be loaded using any tool that supports these common formats. When loading the Parquet, \texttt{int32} will be used automatically.
\begin{lstlisting}
df_cites = pd.read_parquet(
    "citations.parquet",
    engine="pyarrow",
    dtype_backend="pyarrow",
)
\end{lstlisting}

\begin{table}[h]
\centering
\begin{tabular}{l S S}
\hline
\textbf{Dataset} & \textbf{Size on disk (GB)} & \textbf{Memory usage (GB)} \\
\hline
citations.tsv & 39 & 17 \\
publications.tsv & 6 & 5 \\
publications\_large.tsv & 187 & 185 \\
citations.parquet & 8 & 17 \\
publications.parquet & 2 & 5 \\
publications\_large.parquet & 68 & 185 \\
Full OA -- edges & 1820 & {NA} \\
Full OA -- nodes & 700 & {NA} \\
\hline
\end{tabular}
\caption{Comparison of storage size and memory requirements of the full OpenAIRE (OA) dataset (release 2025-12-01) and its compact versions. Memory usage corresponds to the amount of GB each dataset occupied when loaded into a Pandas dataframe \citep{reback2020pandas}. Since the citations are loaded as int32, the memory size is much lower than disk size.}
\label{tab:datasetcomparison}
\end{table}

The publications files benefit from being loaded using the PyArrow backend. This significantly reduces the memory usage of the doi field.
\begin{lstlisting}
df_pubs = pd.read_parquet(
    "publications.parquet",
    engine="pyarrow",
    dtype_backend="pyarrow",
)
\end{lstlisting}

The publications\_large files are the most important to load efficiently as $\sim200$GB of RAM can be saved. Be aware that even efficient loading still requires $\sim185$GB of RAM. For users interested in a subset of the provided columns, we recommend selecting the columns on load. For loading the full dataset the following code can be used, and selected columns can be removed. This efficient loading is the approach used for the "Pandas Opt" column in Table \ref{tab:column_sizes_python_types_desc_left}. In the example below, the columns nodeId, title and pid\_dois are selected.
\begin{lstlisting}
df_large = pd.read_parquet(
    "publications_large.parquet",
    columns=["nodeId", "title", "pid_dois"],
    engine="pyarrow",
    dtype_backend="pyarrow",
)
df_large[["language", "container"]] = (
    df_large[["language", "container"]].astype("category")
)
\end{lstlisting}

\begin{table}[h]
\centering
\begin{tabular}{lllrrr}
\hline
\textbf{Column} & \textbf{Python Type} & \textbf{Description} & \textbf{Arrow} & \textbf{Pandas} & \textbf{Pandas Opt} \\
\hline
nodeId          & int32 & Node ID & 0.767 GB   & 0.767 GB   & 0.767 GB \\
openaireId      & str & OpenAIRE unique ID & 9.585 GB   & 19.746 GB  & 9.586 GB \\
title           & str & Paper title & 16.486 GB  & 29.707 GB  & 16.486 GB \\
authors         & list[str] & List of authors & 11.037 GB  & 23.005 GB  & 11.038 GB \\
description     & str & Main text/abstract & 131.248 GB & 193.583 GB & 131.248 GB \\
date            & datetime & Publication date & 0.767 GB   & 7.587 GB   & 0.768 GB \\
container       & str & Journal/conference name & 5.471 GB   & 13.953 GB  & 2.181 GB \\
citations       & int & Citation count & 1.558 GB   & 1.534 GB   & 1.558 GB \\
language        & str & Language & 1.394 GB   & 11.555 GB  & 0.197 GB \\
pid\_dois        & list[str] & DOI identifiers & 5.639 GB   & 19.453 GB  & 5.639 GB \\
pid\_mag\_ids     & list[str] & MAG IDs & 2.004 GB   & 12.748 GB  & 2.005 GB \\
pid\_pmids       & list[str] & PubMed IDs & 1.202 GB   & 7.948 GB   & 1.202 GB \\
pid\_handles     & list[str] & Persistent handles & 1.149 GB   & 6.134 GB   & 1.149 GB \\
pid\_pmcs        & list[str] & PubMed Central IDs & 0.921 GB   & 5.480 GB   & 0.921 GB \\
pid\_arxiv\_ids   & list[str] & ArXiv IDs & 0.885 GB   & 4.856 GB   & 0.886 GB \\
\hline
\textbf{TOTAL}  & & & 190.113 GB & 358.055 GB & 185.631 GB \\
\hline
\end{tabular}
\caption{Memory size of columns with Python types and short descriptions (Arrow, Pandas, and optimized Pandas formats). Arrow is the memory size loaded using PyArrow, Pandas is the size if the data is loaded straight into a default Pandas dataframe. The Pandas Optimized column uses PyArrow as a backend, and utilizes "categorical" on the container and language fields to further lower their footprint. The  MAG IDs are Microsoft Academic Graph IDs.}
\label{tab:column_sizes_python_types_desc_left}
\end{table}

\section{Dataset Description}
\paragraph{Repository name} Zenodo
\paragraph{Object name} Compact representation of the OpenAIRE citation graph
\paragraph{Format names and versions} TSV (Tab Separated Values) and Apache Parquet
\paragraph{Creation dates} Based on the 2025-12-01 OpenAIRE dump.
\paragraph{Dataset creators} Joakim Skarding wrote the pipeline that processed the dump, Pavel Sanda supervised the work. The OpenAIRE initiative, with which this project is unaffiliated, created the OpenAIRE dump.
\paragraph{Language} English
\paragraph{License} CC BY 4.0
\paragraph{Publication date} 2026-02-12

\section{Reuse Potential}
Citation networks are routinely used in a wide range of scientific fields. This includes among others, research in historical trends of science \citep{frank2019evolution, drivas2024evolution, kitajima2025altering,gonzalez2024landscape}, sociology of scientific knowledge \citep{crothers2020citation,carradore2022academic}, network science \citep{costa2024complexity,xiao2025characterizing} and as training sets for graph neural networks \citep{kipf2016semi,leskovec2016snap}. Since the dataset is an evolving network, it can also be used to train temporal models, such as dynamic graph neural networks \citep{skarding2021foundations}. 

Compared to working directly with the original OpenAIRE data dump, this dataset significantly reduces the time and effort required to begin analysis. The data is provided as flat TSV and Parquet files, removing the need to parse and process the raw JSON. The citation graph is already structured as an edge list with integer node IDs, the format expected by most graph libraries, meaning graph algorithms (e.g. community detection~\citep{fortunato2010community} and centrality measures~\citep{bloch2023centrality}) can be applied directly without further transformation. The reduced file size allows the full dataset to be downloaded and explored locally. Together, these properties make the dataset a convenient starting point for bibliometric studies, network analysis, and graph learning research.

We provide the source code used to produce the data, allowing researchers to run the pipeline when new versions of the OpenAIRE graph is released, as well as make their own customized distilled dataset with the fields they desire.

%\section*{Acknowledgements}
%Please add any relevant acknowledgements to anyone else that assisted with the project in which the data was created but did not work %directly on the data itself.

\section*{Funding Statement}
This work has been funded by a grant from the Programme Johannes Amos Comenius under the Ministry of Education, Youth and Sports of the Czech Republic, CZ.02.01.01/00/23\_025/0008711.

\section*{Competing interests} 
The authors have no competing interests to declare.

%\section*{Supplementary Files (optional)}
%Any supplementary/additional files that should link to the main publication must be listed, with a corresponding number, title and option description. Supplementary files should also be cited in the main text.
%Note: supplementary files will not be typeset so they must be provided in their final form. They will be assigned a DOI and linked to from the publication.

\section*{Supplementary information}

\paragraph{Example of reduction of a relation (citation/edge)}
Our dump has the benefit of focusing solely on citations, and thus we do not need to indicate the citation type. We only retain the source and target IDs. These IDs are also optimized to take up as little space as possible.
The following is an example of how an edge is represented in the OpenAIRE JSON dump:
\begin{verbatim}
{
  "provenance": {
    "provenance": "Inferred by OpenAIRE",
    "trust": "0.9"
  },
  "relType": {
    "name": "Cites",
    "type": "citation"
  },
  "source": "doi_________::7e8d84fc096936557defb78d22cca97c",
  "sourceType": "product",
  "target": "dedup_wf_002::27d83ddfd6e54378d88445aa793d5cb8",
  "targetType": "product",
  "validated": false
}
\end{verbatim}

\noindent An edge in our dump is simply:
\begin{verbatim}
159486578   118392581
\end{verbatim}

\paragraph{Estimates of OpenAIRE JSON files uncompressed}
Table \ref{tab:estimate} shows a rough estimate of the full uncompressed size of the OpenAIRE dump. The dump would require several orders of magnitude more RAM than most consumer computers.

\begin{table}[h!]
\centering
\caption{Estimate of OpenAIRE JSON files when uncompressed}
\begin{tabular}{l r}
\hline
\textbf{Relation (edges) Size Estimate} & \\
\hline
One relation JSON (uncompressed) & 6.5\,GB \\
One relation JSON (compressed) & 0.5\,GB \\
Compression ratio & $6.5 / 0.5 = 13$ \\
Compressed relation folder size & 10\,GB \\
Estimated uncompressed per folder & $13 \times 10 = 130\,\text{GB}$ \\
Number of relation folders & 14 \\
\textbf{Total uncompressed (relations)} & $14 \times 130 = \textbf{1.82\,TB}$ \\
\hline
\textbf{Publication (nodes) Size Estimate} & \\
\hline
One publication JSON (uncompressed) & 0.5\,GB \\
One publication JSON (compressed) & 0.1\,GB \\
Compression ratio & $0.5 / 0.1 = 5$ \\
Compressed publication folder size & 10\,GB \\
Estimated uncompressed per folder & $5 \times 10 = 50\,\text{GB}$ \\
Number of publication folders & 14 \\
\textbf{Total uncompressed (publications)} & $14 \times 50 = \textbf{0.70\,TB}$ \\
\hline
\textbf{Combined Uncompressed Estimate} & \textbf{2.5\,TB} \\
\hline
\label{tab:estimate}
\end{tabular}
\end{table}

\paragraph{Filtered fields}
We remove a lot of fields from the original OpenAIRE graph dump. Table \ref{tab:filtered_fields} is an overview of the fields we have filtered out of the publications (nodes). Table \ref{tab:relation_fields} is an overview of the fields filtered out of the relations (edges).

\begin{table}[ht]
\centering
\small
\begin{tabular}{ll}
\hline
\textbf{Field} & \textbf{Description} \\
\hline
\multicolumn{2}{l}{\textit{Bibliographic}} \\
\texttt{subTitle}            & Alternative or explanatory title \\
\texttt{publisher}           & Publishing entity \\
\texttt{version}             & Version of the result \\
\texttt{sources}             & Dublin Core dc:source \\
\texttt{contributors}        & Contributing persons or entities \\
\texttt{subjects}            & Keywords with scheme and provenance \\
\texttt{formats}             & File formats \\
\texttt{coverages}           & Coverage information \\
\texttt{originalIds}         & Identifiers at original sources \\
\texttt{dateOfCollection}    & When OpenAIRE last collected the record \\
\texttt{lastUpdateTimeStamp} & Timestamp of last update in OpenAIRE \\
\texttt{embargoEndDate}      & Date embargo ends \\
\hline
\multicolumn{2}{l}{\textit{Open Access}} \\
\texttt{bestAccessRight}     & Openest access right (code, label, scheme) \\
\texttt{isGreen}             & True if green Open Access \\
\texttt{isInDiamondJournal}  & True if published in a Diamond Journal \\
\texttt{openAccessColor}     & gold, hybrid, or bronze \\
\texttt{publiclyFunded}      & True if outcome of a funded project \\
\hline
\multicolumn{2}{l}{\textit{Impact indicators}} \\
\texttt{indicators.citationImpact.citationClass}   & Citation class label \\
\texttt{indicators.citationImpact.impulse}         & Impulse score \\
\texttt{indicators.citationImpact.impulseClass}    & Impulse class label \\
\texttt{indicators.citationImpact.influence}       & Influence score \\
\texttt{indicators.citationImpact.influenceClass}  & Influence class label \\
\texttt{indicators.citationImpact.popularity}      & Popularity score \\
\texttt{indicators.citationImpact.popularityClass} & Popularity class label \\
\texttt{indicators.usageCounts.downloads}          & Download count \\
\texttt{indicators.usageCounts.views}              & View count \\
\hline
\multicolumn{2}{l}{\textit{Instances (per-version metadata)}} \\
\texttt{instances}           & Per-version access rights, URLs, license, refereed status \\
\hline
\multicolumn{2}{l}{\textit{Geolocation}} \\
\texttt{geoLocations}        & Box, place, and point geolocation data \\
\texttt{countries}           & Associated countries with ISO codes \\
\hline
\multicolumn{2}{l}{\textit{Software/dataset specific}} \\
\texttt{codeRepositoryUrl}   & URL to source code repository \\
\texttt{contactGroups}       & Groups responsible for the software \\
\texttt{contactPeople}       & Persons responsible for the software \\
\texttt{documentationUrls}   & URLs to software documentation \\
\texttt{programmingLanguage} & Programming language of software \\
\texttt{size}                & Declared size of dataset \\
\texttt{tools}               & Tools for interpretation of result \\
\hline
\multicolumn{2}{l}{\textit{Sub-fields of included objects (partially used)}} \\
\texttt{authors.name}        & Author first name \\
\texttt{authors.surname}     & Author surname \\
\texttt{authors.rank}        & Author position \\
\texttt{authors.pid}         & Author ORCID identifier \\
\texttt{container.edition}   & Journal or proceeding edition \\
\texttt{container.ep}        & End page \\
\texttt{container.iss}       & Journal issue number \\
\texttt{container.issnLinking} & Linking ISSN \\
\texttt{container.issnOnline}  & Online ISSN \\
\texttt{container.issnPrinted} & Printed ISSN \\
\texttt{container.sp}        & Start page \\
\texttt{container.vol}       & Volume \\
\texttt{container.conferenceDate}  & Conference date \\
\texttt{container.conferencePlace} & Conference location \\
\texttt{language.label}      & Language label in English \\
\hline
\end{tabular}
\caption{Fields present in the OpenAIRE product schema that are not included in the output dataset.}
\label{tab:filtered_fields}
\end{table}

\begin{table}[ht]
\centering
\small
\begin{tabular}{ll}
\hline
\textbf{Field} & \textbf{Description} \\
\hline
\multicolumn{2}{l}{\textit{Core relation fields}} \\
\texttt{sourceType}    & Type of the source entity (e.g. product, project, organization) \\
\texttt{targetType}    & Type of the target entity \\
\texttt{relClass}      & High-level category of the relation (e.g. citation, affiliation) \\
\texttt{relType}       & Specific relation type describing the semantic meaning \\
\hline
\multicolumn{2}{l}{\textit{Provenance and validation}} \\
\texttt{provenance}    & Information about how the relation was generated or provided \\
\texttt{validated}     & Indicates whether the relation has been validated \\
\texttt{validationDate} & Date when the relation was validated \\
\hline
\multicolumn{2}{l}{\textit{Additional metadata}} \\
\texttt{confidence}    & Confidence score for the relation \\
\texttt{inference}     & Indicates whether the relation was inferred or explicitly provided \\
\hline
\end{tabular}
\caption{Fields present in the OpenAIRE relation schema that are not included in the output dataset. The core relation field "relType" is used for filtering to only obtain "Cites" relations. And removal of dangling edges ensures that the output dataset only includes edges with sourceType and targetType "product".}
\label{tab:relation_fields}
\end{table}

\end{document}